\documentclass[twocolumn,superscriptaddress,showpacs,prb]{revtex4}
\usepackage{amssymb}
\usepackage{amsmath}
\usepackage{graphicx}

\begin{document}

\title{QSGW calculation of the work functions of Al(111), Al(100), and Al(110) surfaces}
\author{Sergey V. Faleev}
\email{sfaleev@mint.ua.edu}
\author{Oleg N. Mryasov}
\affiliation{MINT Center, University of Alabama, P.O. Box 870209, Tuscaloosa, AL 35487, USA}
\affiliation{Physics and Astronomy, University of Alabama, Tuscaloosa, AL, 35487, USA}

\author{Thomas R. Mattsson}
\affiliation{HEDP Theory, MS 1189, Sandia National Laboratories, Albuquerque, NM 87185, USA}

\date{\today}

\begin{abstract}
Modifications to the quasiparticle self-consistent GW (QSGW) method needed to correctly describe metal/vacuum interfaces and other systems having extended regions with small electron density are identified and implemented.
The method's accuracy is investigated by calculating work functions for the Al(111), Al(100), and Al(110) surfaces.
We find that the results for work function do not depend on the DFT functional employed to calculate the starting
Hamiltonian and that QSGW yield results in quantitative agreement with data from ultrahigh vacuum experiments.
\end{abstract}
\maketitle

%\special{papersize=8.5 in, 11 in}
\section{Introduction}

The work function is not only a most important quantity characterizing the surface of a metal,
it also directly affect surface phenomena like growth rate, the form of crystallites,
sintering, catalytic behavior, adsorption, chemical reactions,
surface segregation, and formation of grain boundaries.
In addition, the work function largely determines rates of electron
surface emission and is therefore of applied interest for optimizing
thermionic emitters~\cite{JenniSS2004}, where a low work function is
sought, and pulsed-power components, e.g, in transmission lines, where
a high work function is required.
An ability to make accurate theoretical calculations of
work functions is therefore of great interest and importance for the development
of new high-performing materials.

Presently, most calculations of work functions are made within
density functional theory (DFT) \cite{Hohenberg64,Kohn65} using either the
local density approximation (LDA) or the generalized gradient approximation (GGA).
Unfortunately, the accuracy of the DFT-based methods in calculation of the work
function is not always satisfactory and results depend on
the DFT functional used (see, e.g., recent review of available theoretical and experimental
values of work function for a number of metals in Ref. \onlinecite{Kawano08}). For instance,
the work functions of Al surfaces calculated by using the GGA with Perdew-Burke-Ernzerhof (GGA/BPE)
functional \cite{PBE96} are approximately 0.2 eV lower then corresponding LDA values calculated
with Ceperley-Alder (LDA/CA) functional  \cite{Ceperley80} and approximately 0.3 eV lower then
corresponding LDA values calculated with Barth-Hedin (LDA/BH) functional \cite{Barth72} (see Table I).
This example highlights the need for methods capable of reaching 0.1 eV (or better) accuracy for
theoretical prediction of the work function of metals.

%could be required for accurate
%(of the order of 0.1eV or better) prediction of the work function.
% (see, e.g., recent review of calculated and experimental
%values of work functions for a number of metals in Ref. \onlinecite{Kawano08} and also
%Table I below). For instance, recent GGA calculations predict the values of the of the work
%function for the Al(111), Al(110), and Al(110) equal to 4.06eV, 4.24eV, and 4.07 eV \cite{Silva05},
%that is about 0.2eV lower then corresponding experimental values of 4.24eV, 4.41eV, and 4.28eV
%obtained by photoelectric measurement carried out in ultrahigh vacuum conditions\cite{Grepstad76}.

The GW approximation of Hedin \cite{Hedin69} is a well established method which
yields highly accurate quasiparticle (QP) energies for bulk materials
\cite{Hybertsen85,Schilfgaarde06}. GW calculations are usually performed in a
non-self-consistent manner by using the LDA Green's function, G, and screened Coulomb interaction,
W (so-called $G_{0}W_{0}$ method). The results obtained by the $G_{0}W_{0}$ method depend on
the quality of underlying DFT wave-functions and eigenvalues and the agreement with experimental data worsen if
the DFT description is not sufficiently accurate. For example, $G_{0}W_{0}$ fails to describe the band
gap of NiO \cite{Faleev04}. Recently, Faleev, van Schilfgaarde
and Kotani \cite{Faleev04,Schilfgaarde06} developed the so-called Quasiparticle Self-consistent
GW (QSGW) method which is independent of DFT and demonstrated
that results for QP energy levels of bulk materials obtained by the QSGW are in better agreement
with experiment then results obtained by the standard $G_{0}W_{0}$ method.  In contrast to the $G_{0}W_{0}$
method, QSGW describes correctly also strongly correlated materials like NiO or MnO \cite{Faleev04}.

Although calculations of surface QP energies were performed already more then two decades ago
\cite{Hybertsen88}, GW calculations for surfaces and other non-bulk systems remains rare due to
the demanding computational requirements. Most commonly, the GW method has been applied to study
an image potential and corresponding image states of insulating \cite{Rohlfing03} and metallic
\cite{White98,Fratesi03,Crampin05} surfaces and clusters \cite{Rinke04}. Recently, the GW method
was used to investigate the image potential-induced renormalization of the molecular
electronic levels for molecules adsorbed on metal surfaces \cite{Neaton06,Thygesen09} and thin insulator
films \cite{Freysoldt09}. Note that the image potential cannot be obtained by the LDA or GGA approaches
since they do not include non-local polarization effects that are present
in GW theory through the W operator. Another application of the GW method to non-bulk materials that is becoming
an active area of research is QP calculations of transport properties of nanoscale systems
\cite{Darancet07,Thygesen07,Spataru09}.
%For instance, recent GW simulations of a gold chain showed that the
%transmission function is considerably modified with respect to the DFT one and compares favorably with trends
%experimentally observed for this system.

To the best of our knowledge, two GW studies of the work functions of metals have been
published to date. Morris et al \cite{Morris07} calculated
the work functions for Al(111), Al(100), and Al(110) surfaces using the $G_{0}W_{0}$ method and making a jellium
approximation. They also included vertex corrections to the self-energy and W (evaluated on a homogeneous
electron gas level) that resulted in a significant, over 1 eV, underestimation of
calculated work functions, attributed to inherent self-interaction error.
% arising when the vertex correction are treated on a homogeneous electron gas level.
%Another study of the Al work function on a beyond LDA level has been performed by
Heinrichsmeier et al \cite{Hein98} proposed a new non-local parametrization of the exchange-correlation
functional derived from the $G_{0}W_{0}$ calculations for jellium surfaces (the
V$_{xc}$(GW) method), and applied the method to real (111) and (100) surfaces of Al and Pt.
A conclusion of both studies was that the values of the Al(111) and Al(100)
work functions  obtained by the $G_0W_0$ method are significantly worse
then corresponding LDA values as compared to the experimental data (see Table I).
Since both these GW calculations involve the jellium approximation,
the question remains how well the fully atomistic GW method could describe the work
functions of metals as compared to the DFT results and experiment.

%  Heinrichsmeier discussion:
%They obtained the values of the work function 4.82 eV for Al(111) and 4.59 eV for Al(100), that
%are 0.58 eV and 0.18 eV larger then experimental results, correspondingly. Such large error,
%particularly for the Al(111), could be attributed to the fact that these GW calculations were
%non-self-consistent; consequently, results  are  dependent on the initial DFT functional parametrization \cite{Hein98}.

% Morris discussion
%They obtained the $G_{0}W_{0}$ values of Al(111), Al(100), and Al(110) work functions that are +0.36 eV, +0.28 eV,
%and +0.02 eV larger then corresponding experimental values of 4.24 eV, 4.41 eV, and 4.28 eV obtained by photoelectric
%measurement carried out in ultrahigh vacuum conditions\cite{Grepstad76}. These differences should be compared with
%LDA differences of -0.06 eV, -0.14 eV, and -0.40 eV.
%
%Morris et al \cite{Morris07} also studied the role of vertex corrections to $G_{0}W_{0}$, treating the vertex
%corrections on a homogeneous electron gas level. They found that if the vertex corrections were included only
%to W, the values of the work functions for all three Al surfaces increase by 0.17eV compare to $G_{0}W_{0}$ (
%thus further modestly worsening the agreement with the experiment); while inclusion of the vertex corrections
%in both W and self-energy $\Sigma$ (which is more theoretically consistent way) resulted in hugely underestimated
%values of work functions: 1.11eV, 1.45eV, and 1.19 eV smaller then experimental values for Al(111), Al(100), and
%Al(110) surfaces correspondingly.

In this paper, we present results of work function calculations for the
Al(111), Al(100), and Al(110) surfaces evaluated within fully atomistic $G_{0}W_{0}$ and QSGW approaches.
As we show below, the values of the work function obtained by these two methods
only insignificantly deviate from each other (within 0.02 eV), do not depend on the initial DFT
functional, and are in excellent agreement with experimental data.
On the other hand, the DFT approaches predict different results
for the work function depending on the functional used.
The paper is organized into two main sections describing the methods and the results
followed by a concise summary.

\section{Method and computational details}

Let us first briefly describe the computational approach.
QSGW is a method to determine nonlocal (but static and Hermitian)
optimum one-particle Hamiltonian $H^{0}$ in a self-consistent way
\cite{Faleev04,Schilfgaarde06,Kotani07}. First, starting with a
trial Hamiltonian $H^{0}$ (usually, the LDA Hamiltonian is used as
the first-iteration $H^{0}$) the self-energy $\Sigma(\omega)$ is calculated
in the GW approximation. The static
self-energy is defined in the basis of the eigenfunctions
$\psi_{\mathbf{k}n}(\textbf{r})$ of the Hamiltonian $H^{0}$ as
follows \cite{Faleev04,Schilfgaarde06,Kotani07}
\begin{equation}
\Sigma^{\mathbf{k}}_{nn'}=Re\langle \psi_{\mathbf{k}n}|[ \Sigma(\varepsilon_{\mathbf{k}n}) +
\Sigma(\varepsilon_{\mathbf{k}n'})]/2 | \psi_{\mathbf{k}n'}\rangle \ , \label{eq1}
\end{equation}
where $\mathbf{k}$ is the wave vector, $n$ is the band index, $\varepsilon_{\mathbf{k}n}$ denote
eigenvalues of $H^{0}$, and $Re$ means to take the Hermitian part. Next, the Hamiltonian
$H^{0}$ is updated for each iteration using $\Sigma^{\textbf{k}}_{nn'}$ instead of the usual LDA
exchange-correlation potential %$V^{xc,LDA}$.
\begin{equation}
H^0=-\frac{\nabla^2}{2m} + V^{ext} + V^H + \sum_{\mathbf{k}nn'}
| \psi_{\mathbf{k}n}\rangle \ \Sigma^{\mathbf{k}}_{nn'} \langle \psi_{\mathbf{k}n'}|  \ , \label{eq2}
\end{equation}
where $V^{ext}$ is the external (nuclei) potential and $V^H$ is the
Hartree potential. This procedure is iterated until
self-consistency is reached. Here, self-consistency is defined as when $\Sigma^{\mathbf{k}}_{nn'}$
generated by $H^{0}$ is identical (within a small tolerance) to
the $\Sigma^{\mathbf{k}}_{nn'}$ that enters into $H^{0}$. It has
been shown \cite{Schilfgaarde06} that this procedure in an approximate way
minimizes the difference between the full
nonlocal, non-static and non-Hermitian GW Hamiltonian
$H(\omega)=-\frac{\nabla^2}{2m} + V^{ext} + V^H + \Sigma(\omega)$
and Hamiltonian $H^{0}$ (that is why it is called 'optimum').  Note
that $H(\omega)$ is a functional of $H^{0}$ because both $V^H$
and $\Sigma(\omega)$ calculated in the GW approximation depend on
eigenfunctions generated by $H^{0}$. Hence, the iteration
procedure described above self-consistently determines both $H(\omega)$ and the
corresponding optimum $H^{0}$.  The $G_{0}W_{0}$ method is simply the first
iteration of described above cycle, self-energy  is obtained from Eq. (\ref{eq1})
using LDA wave function and energies, and  first-iteration $G_{0}W_{0}$ Hamiltonian
is constructed by Eq. (\ref{eq2}).

%\subsection{Computational details}

In the present work, we used the experimental lattice constant of Al at zero
temperature, 4.025 {\AA} \cite{Wyckoff63}. The Al(111), Al(100), and Al(110)
surfaces were modeled by a ($1\times 1$) surface unit cell  in
$xy$-directions and a periodic combination of $N_{A}$ Al layers and
$N_{V}$ vacuum layers in $z$-direction. The vacuum layer was of the
same width as the Al layer, and contains so-called floating basis orbitals
\cite{Schilfgaarde06PRB} placed instead of the atomic muffin-tin
orbitals. $N_{A}$ ranged from 4 to 12 and $N_{V}$ ranged from 6 to
10 were used to analyze the convergence of the results with
respect to these parameters. The work function, $\Phi$, is defined
as the difference between the electrostatic potential at a point far
from the surface and the Fermi energy;
$\Phi=V_{es}(\infty)-\varepsilon_F$. In our calculations
$V_{es}(\infty)$ was estimated as electrostatic potential in the
middle of the vacuum slab.

It is known that GW calculations of band gaps in semiconductor thin films\cite{Freysoldt08,Ismail06} and molecular
chains\cite{Rozzi06}  performed in repeated-cell geometries converge slowly with vacuum thickness due to the
long-range nature of the non-local screened Coulomb interaction.
Rozzi et al. \cite{Rozzi06} investigated this problem and
developed a Coulomb cut-off scheme to eliminate the long-ranged slab-slab interaction.
They found that the introduced Coulomb cutoff parameter mostly affects delocalized
unoccupied states. On the other hand, both quantities that enter the expression for
the work function: the Fermi energy and electrostatic potential are affected only by
occupied states that are localized inside the metal slab and therefore only weekly depend on
the thickness of the vacuum slab. Consequently, the work function converges quickly
with the number of vacuum layers: we found that $N_V$=6 is sufficient to determine the work
function to 0.005 eV accuracy.

Both relaxed and unrelaxed internal atomic coordinates were utilized  to model the
Al surfaces. The relaxed position of Al layers with $N_{A}\geq 10$ were taken from GGA
calculations of Da Silva \cite{Silva05}. In particular, for Al(111) surface the values
of interlayer expansion/contraction were +1.15\%, -0.05\%, +0.46\%, +0.21\%, -0.05\%
(first number is for the surface layer, last number for fifth layer), for Al(100)
surface +1.59\%, +0.44\%, -0.02\%, -0.68\%, -0.56\%, and for Al(110) surface
-7.18\%, +3.87\%, -2.12\%, +2.04\%, +0.82\% [\onlinecite{Silva05}].

%onm
The QSGW method is implemented as an extension of
the all-electron full-potential linear muffin-tin orbital (LMTO)
program suite. The diagram of the self-energy and density/potential
self-consistency cycles that includes
the  LMTO and GW parts of the code are shown on the Figure 2 of
Ref. [\onlinecite{Kotani07}]. The description of the basis sets and
other details of the LMTO and QSGW implementations can be found in
Refs. [\onlinecite{Kotani07}] and [\onlinecite{Schilfgaarde06PRB}].

%onm
The surface Brillouin zone (BZ) integration in the LMTO
part of the code were performed with ($22\times 22$) and
($24\times 24$) Monkhorst-Pack meshes \cite{Monkhorst76}
($\mathbf{k}^{LMTO}$-mesh). The GW self-energy was calculated
with ($6\times 6$), ($8\times 8$), and ($10\times 10$) meshes in the
surface BZ ($\mathbf{k}^{GW}$-mesh).
The modified offset-$\Gamma$ method designed to treat anisotropic systems was employed
to perform $k$-integration of surface BZ in the GW part of the code
(the method is described in Ref.[\onlinecite{Kotani07}], following the Eq. (53)).
%onm
The GW part of the code, where the self-energy is calculated given
the eigenfunctions and eigenvalues of the $H^0$ generated by the LMTO part, is
significantly more computationally demanding than the LMTO calculation.
In practice, it is computationally prohibitive to
calculate the self-energy on the same fine $\mathbf{k}^{LMTO}$-mesh
required for the LMTO part. Thus, a rather
sophisticated procedure that includes several transformations of the
self-energy $\Sigma^{\mathbf{k}}_{nn'}$ between different basis sets
has been developed to interpolate the self-energy calculated by the
GW part of the program on a coarse $\mathbf{k}^{GW}$-mesh to finer
$\mathbf{k}^{LMTO}$-mesh used by the LMTO part of the
program. \cite{Kotani07}

%onm:
However, matrix elements of the self-energy (\ref{eq1}) between states with high energy
($\geq$ 2 Ry) often cannot be interpolated with sufficient accuracy from the $\mathbf{k}^{GW}$-mesh to the
$\mathbf{k}^{LMTO}$-mesh  \cite{Kotani07}. Note that a fine $\mathbf{k}^{LMTO}$-mesh is
required when describing metals.
The latter is in part due to the long range of the LMTO basis set (e.g. the smallest
eigenvalue of the overlap matrix can be of the order of $10^{-10}$).
In order to overcome this
$\mathbf{k}$-interpolation problem, the high-energy part
($\varepsilon^{LDA}_{\mathbf{k}\tilde{n}},\varepsilon^{LDA}_{\mathbf{k}\tilde{m}}>E_{xccut}$)
of the difference between the self-energy and LDA exchange-correlation
potential,
\begin{equation}
\triangle
V^{xc}_{\tilde{n}\tilde{m}}=\Sigma_{\tilde{n}\tilde{m}} -
V^{xc,LDA}_{\tilde{n}\tilde{m}}
\end{equation}
was substituted with a diagonal matrix with the diagonal elements given by linear function of the
LDA energy, $\triangle
V^{xc}_{\tilde{n}\tilde{n}}=a+b\times\varepsilon^{LDA}_{\mathbf{k}\tilde{n}}$.
Here "$\sim$" over the subscript denotes that the function is
represented in the basis of eigenfunctions
$\psi^{LDA}_{\mathbf{k}\tilde{n}}$ of the LDA Hamiltonian (the LDA
basis) with eigenvalues $\varepsilon^{LDA}_{\mathbf{k}\tilde{n}}$.
The energy cutoff parameter $E_{xccut}$ is typically of the order of
2-3 Ry. The constants $a$ and $b$ are fitted from calculated
$\triangle V^{xc}_{\tilde{n}\tilde{n}}$ at lower energies. The
results for the calculated quasiparticle (QP) energies usually
depend weakly on the cutoff parameter $E_{xccut}$ or constants
$a$ and $b$. More details on the $\mathbf{k}$-interpolation procedure and the
method used to control its accuracy for bulk calculations could be
found in Ref. [\onlinecite{Kotani07}], section II, subsection G.

In the LDA basis the optimum Hamiltonian $H^0(\mathbf{k})$ (\ref{eq2})
reads
\begin{equation}
H^0_{\tilde{n}\tilde{m}}(\mathbf{k})\equiv\langle
\psi^{LDA}_{\mathbf{k}\tilde{n}}|H^0|\psi^{LDA}_{\mathbf{k}\tilde{m}}\rangle=
\varepsilon^{LDA}_{\mathbf{k}\tilde{n}}\delta_{\tilde{n}\tilde{m}} +
\triangle V^{xc}_{\tilde{n}\tilde{m}}(\mathbf{k})  \ . \label{eq5}
\end{equation}
%After described above modification of the matrix $\triangle
After modifications of the matrix $\triangle
V^{xc}_{\tilde{n}\tilde{m}}$ as outlined above, the Hamiltonian
$H^0_{\tilde{n}\tilde{m}}(\mathbf{k})$ in a form of Eq. (\ref{eq5})
is used in the LMTO part of the program to obtain the QP wave functions and energies.

In the present work, we found that the QSGW method requires additional modification when
the system exhibits extended regions with small electronic density, for example
when modeling a metal/vacuum interface.
Explicitly, the matrix elements $\triangle
V^{xc}_{\tilde{n}\tilde{m}}$ in the LDA Hamiltonian of Eq. (\ref{eq5}) leads to
slightly improper mixing between occupied LDA states with energies
$\varepsilon^{LDA}_{\mathbf{k}\tilde{m}}\leq \varepsilon_F$ that are
spatially concentrated in the metal region and "vacuum"~LDA states with energies
$\varepsilon^{LDA}_{\mathbf{k}\tilde{n}}\geq \varepsilon_F + \Phi $
extending to the entire volume of the system
(here $\varepsilon_F$ is the Fermi energy). As a result,
after diagonalization of the Hamiltonian (\ref{eq5}), the QP
occupied states have small tails that decay unphysically slow
as a function of distance from the metal surface.
Several reasons may contribute to this slow, non-exponential decay of occupied QP
wave functions into vacuum, for example: remaining errors in the \textbf{k}-interpolation of
the $\triangle V^{xc}_{\tilde{n}\tilde{m}}$, non-completeness of the LDA basis,
numerical errors, etc.

Development of a general procedure to construct optimum QSGW Hamiltonian for systems with
vacuum regions in a way that \emph{guarantees} correct exponential decay of occupied QP wave
functions in vacuum is beyond the scope of this paper.
However, in application to the specific case of Al, we can overcome this problem by
straightforward truncation of the unphysical non-diagonal matrix
elements $\triangle V^{xc}_{\tilde{n}\tilde{m}}$ with $\varepsilon^{LDA}_{\mathbf{k}\tilde{m}}\leq \varepsilon_F$ and
$\varepsilon^{LDA}_{\mathbf{k}\tilde{n}}\geq \varepsilon_F + \Phi $, using the fact that
the wave functions of bulk Al are rather well described by LDA (see Fig. \ref{fig1}).
Specifically, we truncate  all non-diagonal matrix elements of $\triangle
V^{xc}_{\tilde{n}\tilde{m}}$ if the energies
$\varepsilon^{LDA}_{\mathbf{k}\tilde{n}}$ and
$\varepsilon^{LDA}_{\mathbf{k}\tilde{m}}$ satisfy following
conditions:
\begin{eqnarray}
&&\triangle V^{xc}_{\tilde{n}\tilde{m}}(\mathbf{k})=\triangle
V^{xc}_{\tilde{m}\tilde{n}}(\mathbf{k})=0 \ \ \mbox{ if } \nonumber \\
&&\varepsilon^{LDA}_{\mathbf{k}\tilde{n}}-\varepsilon^{LDA}_{\mathbf{k}\tilde{m}}>E_c
\ \  \mbox{  and   } \ \
\varepsilon^{LDA}_{\mathbf{k}\tilde{n}}>\varepsilon_F \ . \label{eq6}
\end{eqnarray}
Here $E_c$ is a cutoff parameter. The idea of the method is to
allow the \emph{occupied} LDA states $\tilde{m}$ to  be mixed by matrix $\triangle
V^{xc}_{\tilde{n}\tilde{m}}$ only with
\emph{unoccupied} LDA states $\tilde{n}$ with energy \emph{less} then
$\varepsilon_F+E_c$. Such procedure will prevent occupied states from mixing with
extended vacuum states if $E_c<\Phi$ [at least in the first order of
the perturbation theory, if consider $\triangle
V^{xc}_{\tilde{n}\tilde{m}}$ in Eq. (\ref{eq5}) as a
perturbation to the LDA Hamiltonian].
Note that second condition in (\ref{eq6}) always allows the
\emph{occupied} states to be mixed between themselves. In the limit  $E_c \rightarrow
\infty$ there is no modification of the matrix $\triangle
V^{xc}_{\tilde{n}\tilde{m}}$ and the method reduces to
the standard QSGW approach. In the opposite limit, $E_c \rightarrow
+0$, the method becomes an "unoccupied states eigenvalue-only"
self-consistent GW method. The "unoccupied states eigenvalue-only"  means that unoccupied
subblock of the matrix $\triangle V^{xc}_{\tilde{n}\tilde{m}}$ is diagonal, so
the unoccupied QP wave functions are always equal to the LDA wave functions and only QP energies are
modified due to the diagonal matrix elements $\triangle
V^{xc}_{\tilde{n}\tilde{n}}$. Thus, parameter
$0<E_c<\infty$ smoothly interpolates between these two methods.

\section{Results and discussion}

\begin{figure}[t]
\includegraphics*[width=8.5cm]{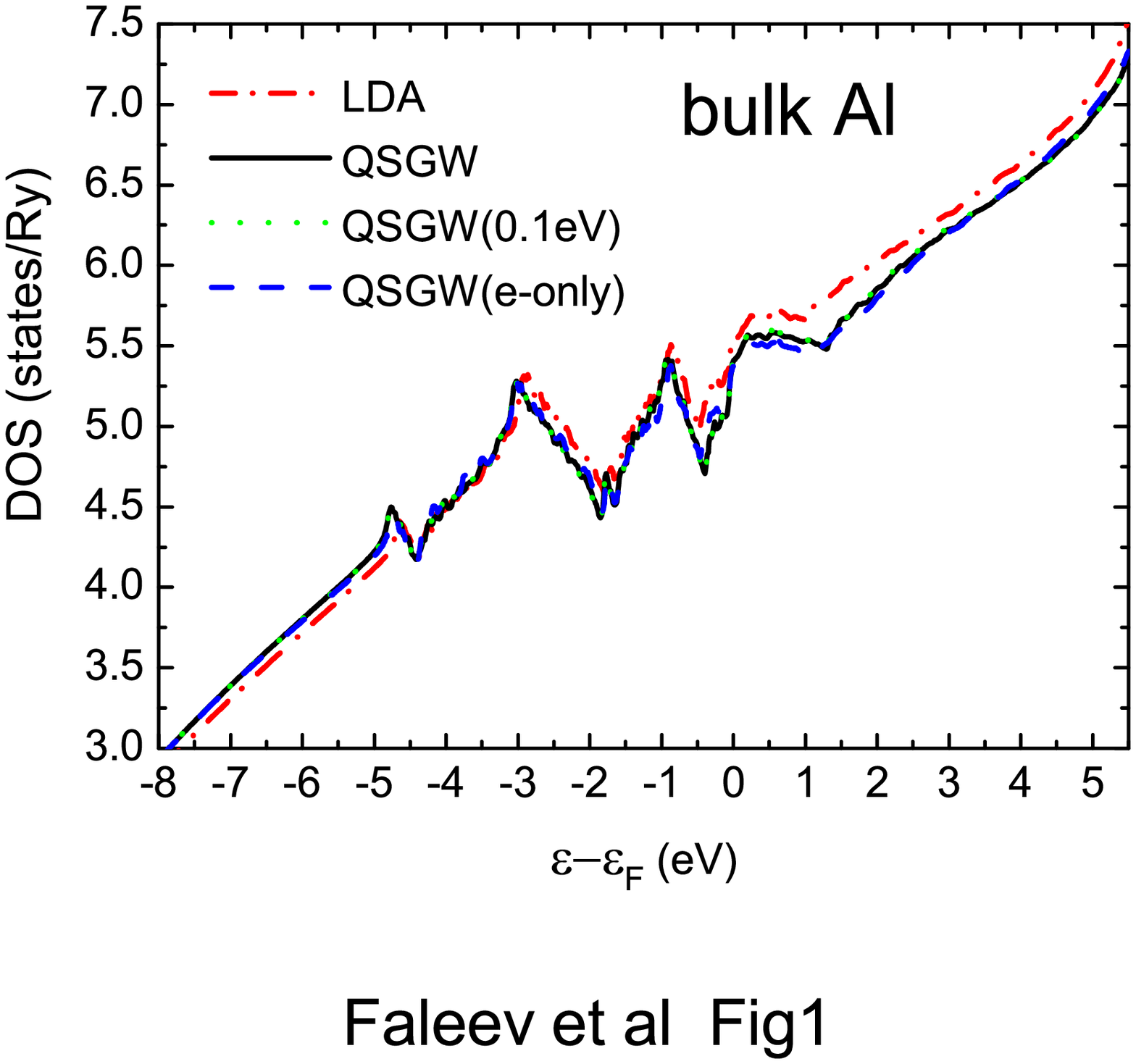}
\caption{(Color online) Density of states (DOS) of
bulk Al calculated in LDA (dot-dashed line), full QSGW that
corresponds to the limit $E_c \rightarrow \infty$ (solid line),
modified QSGW with small cutoff parameter $E_c=0.1$ eV (dotted line), and
the standard "eigenvalue only" self-consistent GW (dashed line). The
DOS curves of the full QSGW ($E_c \rightarrow \infty$) and modified
QSGW with $E_c=0.1$ eV are indistinguishable on the figure.}
\label{fig1}
\end{figure}

We begin with analyzing different characteristics of bulk Al and Al surfaces as function of
parameter $E_c$. Figure~\ref{fig1} shows the density of states (DOS) of the bulk Al
calculated by LDA, full QSGW (that corresponds to the limit $E_c
\rightarrow \infty$), modified QSGW as specified by Eq. (\ref{eq6})
with small cutoff parameter $E_c=0.1$ eV, and the standard
"eigenvalue-only" self-consistent GW methods. (In the "eigenvalue-only" self-consistent GW method
all non-diagonal elements of the GW addition to the LDA Hamiltonian are neglected $\triangle
V^{xc}_{\tilde{n}\tilde{m}}=\triangle
V^{xc}_{\tilde{n}\tilde{n}}\delta_{\tilde{n}\tilde{m}}$, so the QP wave functions are always
equal to the LDA ones.) It is evident that all three modifications
of the GW method produce very similar DOS, somewhat different from the LDA
result. The "eigenvalue-only" DOS is very close to the full QSGW DOS
with minor deviations in the energy range from
$\varepsilon_F-1$ eV to $\varepsilon_F+ 3$ eV. More importantly, the
DOS obtained in the full QSGW (solid line) and DOS obtained by
modified QSGW with small cutoff $E_c=0.1$ eV (dotted line) are
almost indistinguishable on the figure. This means that the
truncation (\ref{eq6}) of the matrix elements  $\triangle
V^{xc}_{\tilde{n}\tilde{m}}$ with \emph{arbitrary}
cutoff parameter $E_c\geq0.1$ eV practically does not change the
\emph{bulk} Al electronic structure. This is an important result suggesting that
we can safely neglect erroneous non-diagonal elements of $\triangle V^{xc}_{\tilde{n}\tilde{m}}$
with $\varepsilon^{LDA}_{\mathbf{k}\tilde{m}}\leq \varepsilon_F$ and
$\varepsilon^{LDA}_{\mathbf{k}\tilde{n}}\geq \varepsilon_F + \Phi $ in \emph{surface}
calculations.

\begin{figure}[t]
\includegraphics*[width=8.5cm]{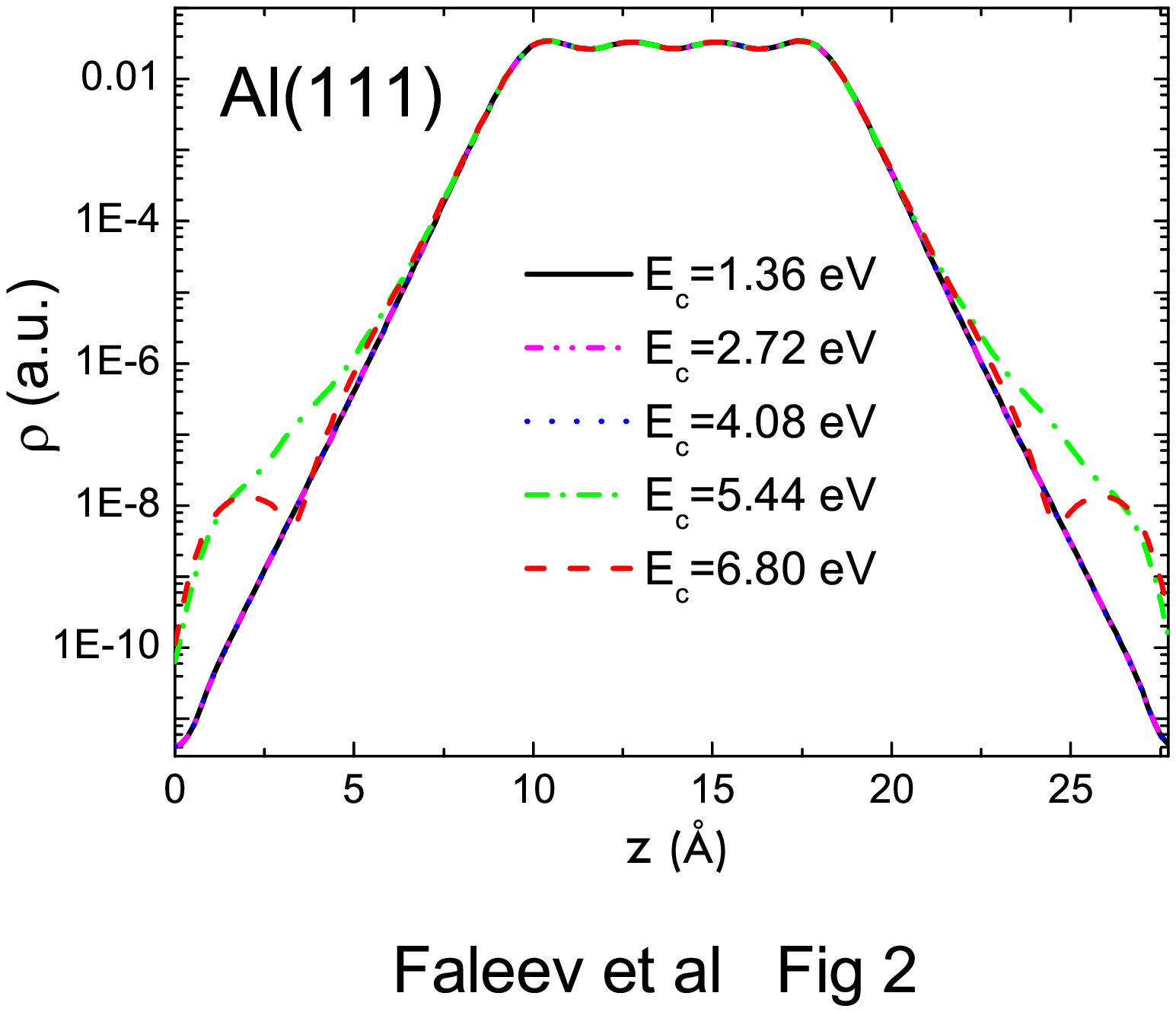}
\caption{(Color online) The electron density averaged over the $xy$  plane
(in atomic units, note  logarithmic scale) as a
function of the distance from the center of the vacuum slab, $z$,
(in {\AA}) for Al(111)/vacuum
interface calculated by modified QSGW method using five
different values of the $E_c$ parameter. The number of Al and vacuum
layers are $N_A=4$ and $N_V=8$.
The density curves with three smallest values of $E_c$
are indistinguishable on the figure.}
\label{fig2}
\end{figure}

Next, we turn to the Al/vacuum interface systems.
Figure~\ref{fig2} shows the electron density averaged over the $x$ and $y$ directions
for an Al(111) surface using five different values of $E_c$.
The calculations were performed for 4 Al and 8 vacuum layers ($N_A=4$ and $N_V=8$).
The expected well behaved exponential decrease of electron density away from the metal surface is seen
for $E_c=1.36$ eV, $2.72$ eV, and $4.08$ eV. Importantly, the densities obtained using these three $E_c$ are
indistinguishable. On the other hand, for $E_c$ above $4.08$ eV, the density begins to deviate from the normal behavior;
it sharply increases near the center of the vacuum.
The density calculated with $E_c=6.8$ eV even increases, at
some $z$, when the distance from the metal increases.
Similar results, independence on $E_c$ and correct exponential behavior of density in vacuum for $E_c \leq 4$ eV, and  unphysical behavior for $E_c \geq 4$ eV, is found for Al(100) and Al(110) surfaces.

We note that the unphysical behavior of the electron density only occur for small absolute density values,
4-5 orders of magnitude smaller then the density in metal region. Also, the QP energy bands depend only weakly on
$E_c$: even the energy bands calculated with parameters $E_c=2.72$ eV and $6.80$ eV
(not shown) almost coincide with each other for states with energies
less then $\varepsilon_F + 2$ eV and begin to deviate slowly for
higher energies. Increasing $E_c$ from $2.72$ eV to $6.8$ eV
results in an up-shift of the QP bands with energy above $\varepsilon_F
+ 4$ eV  by a value ranged from 0 to 0.2 eV, depending on
particular band.

\begin{figure}[t]
\includegraphics*[width=8.5cm]{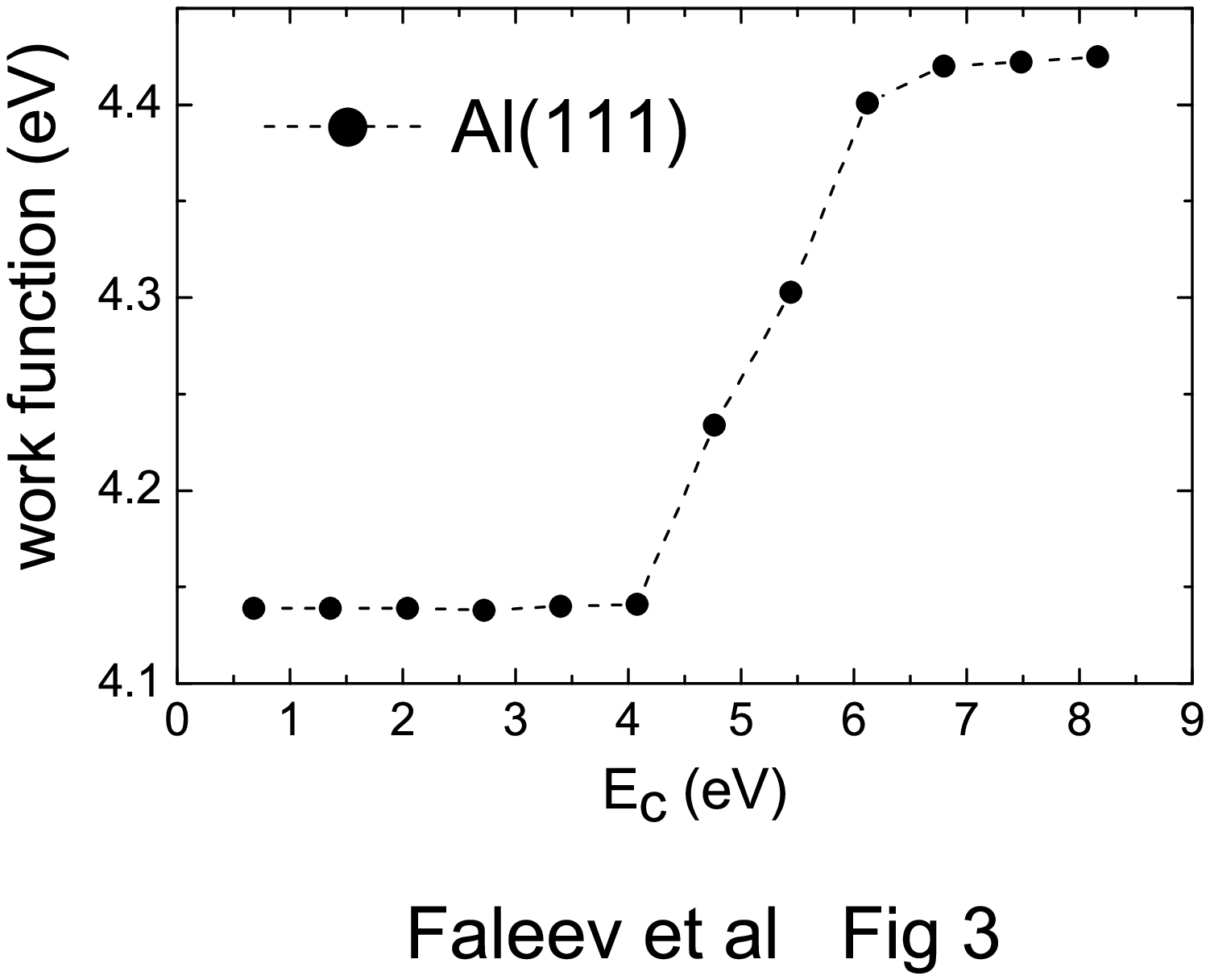}
\caption{ The work function calculated for
Al(111) with $N_A=4$ and $N_V=8$ as function
of the cutoff parameter $E_c$.}
\label{fig3}
\end{figure}

Figure~\ref{fig3} shows the calculated work function, $\Phi$, for Al(111) as function of the
cutoff parameter $E_c$. The value of the work function does not depend on $E_c$, up to $E_c\sim 4$ eV
but changes above this threshold. Similar behavior - independence of the work
function on $E_c$ for $E_c\leq 4$ eV and rapid change above $E_c\sim 4$ eV threshold, was obtained for Al(100) and Al(110)
surfaces. The threshold at $E_c\sim 4$ eV at which point the behavior of the work function and electronic density
in vacuum both sharply changes, is roughly equal to the value of the work function
$E_c\sim \Phi \sim 4.2 - 4.4$ eV of Al surfaces, an additional indication that the origin of
the error is an improper mixing of the occupied and vacuum states
with energies $\varepsilon^{LDA}_{\mathbf{k}\tilde{n}}\geq \varepsilon_F + \Phi $.

Summarizing the results shown in Figures 1-3, we conclude that
(1) The DOS of bulk Al does not depend on $E_c$ for $E_c\geq 0.1$ eV;
(2) for all three Al surfaces the electron density is exponential in vacuum, does not depend on $E_c$ for
$E_c \leq 4$ eV, and demonstrates unphysical behavior for $E_c \geq 4$ eV; and
(3) for all three Al surfaces the value of the work function does not depend on $E_c$ for $E_c \leq 4$ eV,
and sharply changes for $E_c \geq 4$ eV.
Therefore, in the range $0.1$ eV $< E_c < 4$ eV, $E_c$ is large enough for $\triangle V^{xc}_{\tilde{n}\tilde{m}}$ to
include all important matrix elements (at least at the level of bulk Al), and simultaneously
small enough to not include $\triangle V^{xc}_{\tilde{n}\tilde{m}}$ that erroneously mix occupied LDA
states with vacuum LDA states. Importantly, the work function in this range does not depend on
$E_c$ and thus could be taken as the true QSGW value of the work function. Therefore,
in all calculations presented below the $E_c$ parameter is fixed and set to $E_c=2.72$ eV.

The range of applicability of the method described by Eq. (\ref{eq6}) is limited to materials (such as Al)
for which LDA wave functions are adequate so the matrix elements of
$\triangle V^{xc}_{\tilde{n}\tilde{m}}$  with
$\varepsilon^{LDA}_{\mathbf{k}\tilde{n}}-\varepsilon^{LDA}_{\mathbf{k}\tilde{m}}\geq \Phi$ can be neglected.
For any given metal, this condition can be verified on bulk level without performing time consuming surface calculations.
The method is not applicable to metals (such as \emph{d}-electron Fe and Cu) for which the matrix elements
$\triangle V^{xc}_{\tilde{n}\tilde{m}}$ with
$\varepsilon^{LDA}_{\mathbf{k}\tilde{n}}-\varepsilon^{LDA}_{\mathbf{k}\tilde{m}}>\Phi$
play significant roles. As mentioned above, further efforts are required to develop an
universal QSGW-derived method applicable to Fe, Cu and other metals for which simple
truncation of the matrix elements (\ref{eq6}) does not work.

\begin{figure}[t]
\includegraphics*[width=8.5cm]{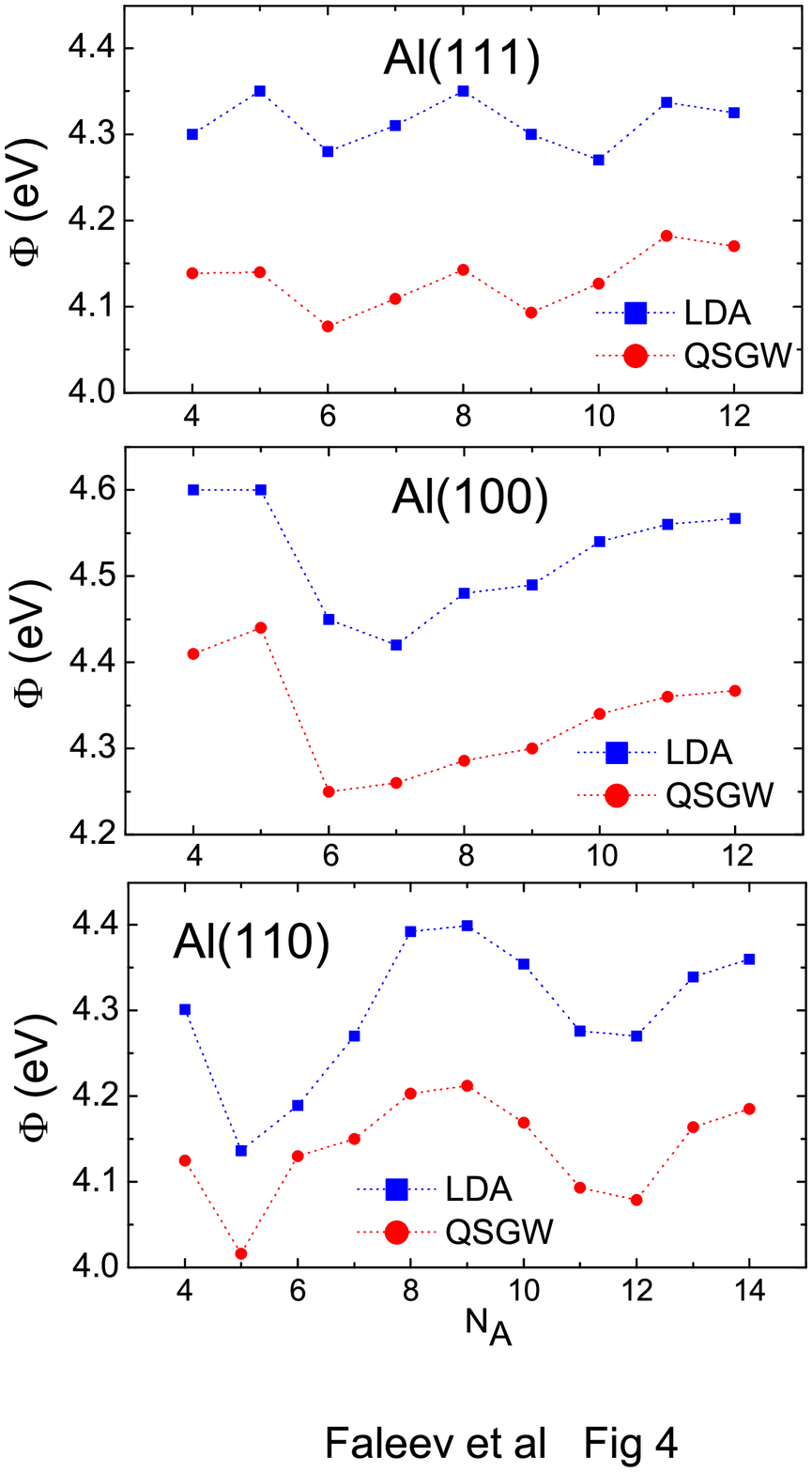}
\caption{ (Color online) The work function of Al(111) (top
panel), Al(100) (middle panel), and Al(110) (bottom panel) surfaces calculated by LDA
(squares) and QSGW (circles) approaches as function of the number of Al layers,
$N_A$.}
\label{fig4}
\end{figure}

Figure~\ref{fig4} shows the variation of the calculated work function, $\Phi$, as function of slab thickness $N_A$.
The calculations were performed with LDA and QSGW for unrelaxed surfaces using the following
parameters: $E_c=2.72$ eV, $N_V=6$, ($22\times 22$)
$\mathbf{k}^{LMTO}$-mesh, and ($6\times 6$) $\mathbf{k}^{GW}$-mesh
in the surface BZ (for the more anisotropic Al(110) surface we used ($22\times 16$)
$\mathbf{k}^{LMTO}$-mesh, and ($6\times 4$) $\mathbf{k}^{GW}$-mesh). For the LDA calculations
we used the Barth-Hedin \cite{Barth72} functional. $\Phi$
oscillates as the slab thickness increases. These oscillations
are well known \cite{Silva05} and can be attributed to quantum-size effects (QSE).
The positions of the local maximums of
the LDA $\Phi$  at $N_A=5$, $8$, and $11$ for the Al(111) surface, and
minimums at $N_A=7$ for the Al(100) surface and at $N_A=5$ and $12$ for the Al(110) surface
are in agreement with previous DFT calculations \cite{Silva05}. The work functions obtained by the QSGW
method show similar QSE oscillations. We estimate the uncertainty in our calculated for
$N_A = 12$ [$N_A = 14$ for Al(110)] values of the work function due to the QSE as $\pm 0.03$ eV,
which is larger then uncertainties due to other computational parameters like the
number of $\mathbf{k}$ points in the $\mathbf{k}^{GW}$-mesh.

Delerue et.\ al \cite{Delerue03} and Freysoldt et.\ al \cite{Freysoldt09} found sizable renormalization of the GW self energy in thin semiconductor films due to the image potential at the interface; this effect
is as large as 0.2 eV for the band gap of Si slabs with a thickness below 3 nm \cite{Delerue03}. This is a finite size effect, different from QSE.  On the other hand, for metallic films the image potential inside
the metal slab is well screened, so it has only a minor effect on occupied states concentrated
within the slab. Since the work function is mostly affected by occupied states, we do not expect a significant image potential induced correction to the value of the QSGW work function. Furthermore, because there is no image potential in the LDA approach, this assumption is supported by the similar behavior of $\Phi$ for QSGW and LDA as a function of slab thickness, see Fig. ~\ref{fig4}.

%
% Ann:
% The results of our calculations of the work function of Al surfaces are shown in Table I in comparison
% with the results of several most recent theoretical studies the experimental data.

Results from our work function calculations for three Al surfaces are shown in Table I in comparison
with experimental data and results from other theoretical studies.
For all three surfaces,  our LDA/CA  results are relatively
close to those obtained by  other groups. The values of work
function of Al(111) and Al(100) surfaces calculated by different groups using the
GGA/BPE method also are relatively close to each other.
Thus, one can conclude that the results obtained using a specific DFT functional
are converged (within 0.1 eV or better accuracy) for different code implementations. On the other hand,
Table I shows that the work functions obtained by using \emph{different} DFT functionals could deviate by
more then 0.1 eV: the LDA/CA values of work functions are universally smaller by $\sim0.1$ eV then LDA/BH values while GGA/PBE values are universally smaller then both LDA/BH and LDA/Wagner values by as much as 0.3 eV. As mentioned in the introduction, such discrepancies emphasize the need for improved methods if an accuracy of 0.1 eV or better is required.

\begin{table}[h]
\caption{A comparison of experimental values of Al work functions (given in eV) for (111), (100),
and (110) surfaces with corresponding values calculated by different methods:
QSGW, $G_0W_0$,  $G_0W_0$ with Al described by jellium ($G_0W_0$[jel]), V$_{xc}$(GW),
LDA/BH, LDA/CA, LDA with Wigner interpolation formula \cite{Wigner34},
and GGA/BPE.}
\centering
%\begin{tabular}{l l l l l l l}
\begin{tabular}{l c c c c c c}
  \hline \hline
  %&\ & \multicolumn{1}{c}{Al(111)} &\ & \multicolumn{1}{c}{Al(100)} \\ [0.5ex]
       &\ & Al(111) &\ & Al(100) &\ & Al(110) \\ [0.5ex]
  \cline{1-1} \cline{3-3} \cline{5-5} \cline{7-7}
  QSGW         &\ &  4.17$^a\ $          \ \ \ &\ &     4.36$^a\ $          \ \ \ &\ &  4.19$^a\ $         \\
  $G_0W_0$     &\ &  4.18$^a\ $          \ \ \ &\ &     4.38$^a\ $          \ \ \ &\ &  4.20$^a\ $         \\
  $G_0W_0$[jel]&\ &  4.60\cite{Morris07} \ \ \ &\ &     4.69\cite{Morris07} \ \ \ &\ &  4.30\cite{Morris07}\\
  V$_{xc}$(GW) &\ &  4.82\cite{Hein98}   \ \ \ &\ &     4.59\cite{Hein98}   \ \ \ &\ &                     \\ [0.5ex]
  LDA/BH       &\ &  4.32$^a\ $          \ \ \ &\ &     4.56$^a\ $          \ \ \ &\ &  4.36$^a\ $         \\ [0.5ex]
  LDA/CA       &\ &  4.22$^a\ $          \ \ \ &\ &     4.46$^a\ $          \ \ \ &\ &  4.26$^a\ $         \\
  LDA/CA       &\ &  4.25\cite{Fall98}   \ \ \ &\ &     4.38\cite{Fall98}   \ \ \ &\ &  4.30\cite{Fall98}  \\
  LDA/CA       &\ &  4.19\cite{Hei93}    \ \ \ &\ &     4.41\cite{Hei93}    \ \ \ &\ &                     \\
  LDA/CA       &\ &  4.21\cite{Silva06}  \ \ \ &\ &                         \ \ \ &\ &                     \\ [0.5ex]
  LDA/Wigner   &\ &  4.31\cite{Sch95}    \ \ \ &\ &     4.51\cite{Sch95}    \ \ \ &\ &   4.32\cite{Sch95}  \\
  GGA/PBE      &\ &  4.06\cite{Silva05}  \ \ \ &\ &     4.24\cite{Silva05}  \ \ \ &\ &  4.07\cite{Silva05} \\
  GGA/PBE      &\ &  4.06\cite{Fall02}   \ \ \ &\ &     4.25\cite{Fall02}   \ \ \ &\ &                     \\
  GGA/PBE      &\ &  4.09\cite{Kj01}     \ \ \ &\ &     4.27\cite{Sferco07} \ \ \ &\ &                     \\ [0.5ex]
  Experiment   &\ &  4.24$\pm$0.02\cite{Grepstad76} \ \ \ &\ &  4.41$\pm$0.03\cite{Grepstad76} \ \ \ &\ & 4.28$\pm$0.02\cite{Grepstad76} \\
  Experiment   &\ &  4.26$\pm$0.03\cite{Eastment73} \ \ \ &\ &  4.20$\pm$0.03\cite{Eastment73} \ \ \ &\ & 4.06$\pm$0.03\cite{Eastment73} \\ [0.5ex]
%  LDA  &\ &\  3.70\cite{Batra86}   \ \ &\  4.59\cite{Boettger96}  \ \ \ &\ &\  3.36\cite{Benesh96}    \ \ &\  4.48\cite{Benesh96}  \\
%  LDA  &\ &\  4.09\cite{Kiejna01}  \ \ &\  4.31\cite{Feibelman84} \ \ \ &\ &\  4.44\cite{Benesh96}    \ \ &\  4.39\cite{Benesh96}  \\
%  LDA  &\ &\  4.54\cite{Skriver92} \ \ &\  4.32\cite{Ho85}        \ \ \ &\ &\  4.82\cite{Benesh96}    \ \ &\  4.53\cite{Wimmer81}  \\
%  LDA  &\ &\  4.49\cite{Hummel98}  \ \ &\  4.7\cite{Krakauer81}   \ \ \ &\ &\  4.50\cite{Hummel98}   \ \ &\   4.51\cite{Bohnen88}  \\ [0.5ex]
  \hline  \hline
  \multicolumn{7}{l}{$^a$Present work}\\
%  \multicolumn{7}{l}{$^b$Present work, Barth-Hedin LDA functional}\\
%  \multicolumn{7}{l}{$^c$Present work, Ceperley-Alder LDA functional} \\

\end{tabular}
\label{tab1}
\end{table}

Table I shows that the work functions calculated using the QSGW method for relaxed Al(111), Al(100),
and Al(110) surfaces are equal to 4.17 eV, 4.36 eV, and 4.19 eV, respectively. We verified that these values
do not depend on the particular LDA exchange-correlation functional used for the initial iteration by applying
both LDA/BH and LDA/CA.
We found that relaxation of the Al surface leads to a less then 0.01 eV shift in the value
of the QSGW work function: work functions for unrelaxed systems are 0.002 eV and 0.007 eV higher for
A(111) and Al (100), and  0.008 eV lower for Al (110) surface relative to corresponding values for relaxed surfaces.
Such small effects of surface relaxation agree well with previous
DFT calculations (see, e.g., Ref. \onlinecite{Silva05}).

When compared with data from experimental photoelectric measurements
carried out under ultrahigh vacuum \cite{Grepstad76}, the work functions obtained using the QSGW method for Al(111), Al(100), and Al(110) surfaces differ by 0.07 eV, 0.05 eV, and 0.09 eV, respectively.
All three differences are less then 0.1 eV and of the order of the sum
of the theoretical and experimental error bars; we therefore consider this agreement excellent.
Of particular interest are the differences between different surface
faces: $\Phi(100)-\Phi(111)=0.19$ eV and $\Phi(110)-\Phi(111)=0.02$ eV are both
in agreement with the experimental data \cite{Grepstad76}
$\Phi(100)-\Phi(111)=0.17$ eV and $\Phi(110)-\Phi(111)=0.04$ eV.

Note that all calculated work functions presented in Table I (except $G_0W_0$[jel]) follow the increasing
trend $\Phi(111) < \Phi(110) < \Phi(100)$, in agreement with data.
This behavior is considered an anomaly; most other fcc metals instead follow Smoluchowski's rule
$\Phi(110) < \Phi(100) < \Phi(111)$. The anomaly is caused by an increased \emph{p}-atomic-like character
of DOS at the Fermi energy in aluminum for the three surfaces, a behavior different to that of
most fcc metals \cite{Fall98}.
We in this context note that earlier experiments \cite{Eastment73} reported 0.21 eV and 0.22 eV
smaller values for Al(100) and Al(110) work functions compare to that of Ref. [\onlinecite{Grepstad76}].
However, Grepstad et al.\cite{Grepstad76} suggested that this discrepancy could be due to
higher impurity concentration, in particular oxygen, in the earlier experiment.

Table I shows that Al work functions calculated using the $G_0W_0$ method differ little
(0.01-0.02 eV) from the converged QSGW results. The $G_0W_0$ results presented in Table I correspond to using LDA/BH as
starting point for GW iterations. Similar 0.01-0.02 eV deviations from the converged QSGW results were found
for  $G_0W_0$ when instead using LDA/CA. We also note that the convergence of the GW iterations is fast, meaning that the initial LDA wave functions are close to the converged QP wave functions; a conclusion supported by
the similarity of the QSGW and QSGW(e-only) DOS for bulk Al shown in Fig. 1.
The substantial differences to previous $G_0W_0$ calculations
by Morris et al \cite{Morris07} ($G_0W_0$[jel] line in Table I) and
Heinrichsmeier et al \cite{Hein98} (V$_{xc}$(GW) line in Table I) should therefore neither
be attributed to the non-self-consistency of the $G_0W_0$ method nor to errors associated with
the choice of particular DFT functional. Instead, we propose that the differences are due to the
jellium approximation employed in both those studies \cite{Morris07,Hein98}.

For more correlated materials such as Fe or Cu
(where LDA and GW wave functions overlap less than they do in Al)
we expect larger deviations of the work functions calculated by $G_0W_0$ and QSGW methods
as well as a stronger dependence of the $G_0W_0$ results on the DFT functional used to calculate $G_0$ and $W_0$.

%  Thomas here

\section{Summary}

We have applied the QSGW and $G_0W_0$ methods to calculate the work functions of
Al(111), Al(100), and Al(110) surfaces. The $G_0W_0$ results
differ from converged QSGW results by less then 0.02 eV and this small difference can be
attributed to significant overlap of the LDA and QP wave functions.
The QSGW results are in excellent agreement with experimental data taken under ultrahigh vacuum conditions.
The calculated values of the work functions do not depend on the DFT functional used for the initial
Hamiltonian $H^0$. These results suggest that QSGW method can be
used for reliable and accurate calculation of the work functions
with accuracy of the order of 0.1 eV or better.

We found that modifications of the original QSGW  method
\cite{Faleev04,Schilfgaarde06,Kotani07} are required in order to
apply the method to the metal/vacuum surface. In particular, special
care should be taken to control the errors originated form (slight) improper mixing
of the occupied and vacuum states.
In some simple cases, such as Al, where LDA wave
functions are already a good approximation to the QP wave functions,
simple truncation of corresponding matrix elements [see Eq.
(\ref{eq6})] are enough to control these errors.

The truncation method is not applicable to metals such as \emph{d}-electron Fe and Cu for which
the matrix elements $\triangle V^{xc}_{\tilde{n}\tilde{m}}$ with
$\varepsilon^{LDA}_{\mathbf{k}\tilde{n}}-\varepsilon^{LDA}_{\mathbf{k}\tilde{m}}>\Phi$
play significant roles. For any given metal, this condition can be verified by studying the bulk electronic structure,
thus without performing time consuming surface calculations. Further efforts are required to develop an \emph{universal} QSGW-derived method applicable to Fe, Cu and other
metals for which simple truncation of the matrix elements is not adequate.

\section{Acknowledgement}
We thank Mark van Schilfgaarde for helpful discussions.
This work was supported by the Science of Extreme Environments
LDRD Investment Area at Sandia National Laboratories.
Sandia is a multiprogram laboratory operated by Sandia
Corporation, a Lockheed Martin Company, for the United States
Department of Energy under contract DE-AC04-94-AL85000.
O.M. and S.F. acknowledge the CNMS User support by Oak Ridge National
Laboratory Division of Scientific User facilities, Office of Basic
Energy Sciences, U.S. Department of Energy.

\end{document}